\documentstyle[epsfig,fullpage,a4]{article}

\renewcommand{\baselinestretch}{1.25}
\begin{document}
\begin{center}

\vspace*{5cm}
{\bf On the solution of the polarisation gain \\ terms for VLBI data
  collected with antennae \\ having Nasmyth or E-W mounts.}

R Dodson

{\bf Sobre la soluci\'on de polarizaci\'on para los datos de VLBI \\
 obtenidos con antenas de sistema \'optico Nasmyth o E-W. \\
    Informe T\'ecnico IT-OAN 2007-16}

\end{center}

\pagebreak

\tableofcontents

\pagebreak

\renewcommand{\thesection}{\Roman{section}:}
\section{Abstract}

I report on the development of new code to support the Nasmyth and
E-W antenna mount types in AIPS which will allow polarisation analysis
of observations made using these uncommon antenna
configurations. These mount types will probably become more widely
spread as they have several advantages, particularly for geodetic
observatories.
Multi-band observations, with multiple receivers, can only be fitted
into telescopes with Nasmyth feeds. These are the requirements for the
new generation of geodetic arrays as discussed in IVS2010. Further
more the next generation of antennae will also be required to have
very high slew rates, and these can be achieved with the E-W mount.

The mount type affects the differential phase between the left and the
right hand circular polarisations (LHC and RHC) for different points
on the sky. The target antennae for the project is the Yebes 40m
telescope, but as that is still under construction the data used as a
demonstration was from the Pico Veleta antenna as part of the Global
Millimeter VLBI Array (GMVA). For the E-W mount type there are
suitable data from the Australian LBA array.
I have demonstrated the effectiveness of the changes made and that the
Nasmyth and E-W corrections can be applied. 

This report does not cover the basics of interferometry, nor regular
polarisation calibration. For references on these topics see, for
example, Thompson, Moran, and Swenson (2001) and Aaron (1997).

\section{Observations of polarisation}

\begin{figure}[htb]
\begin{center}
\epsfig{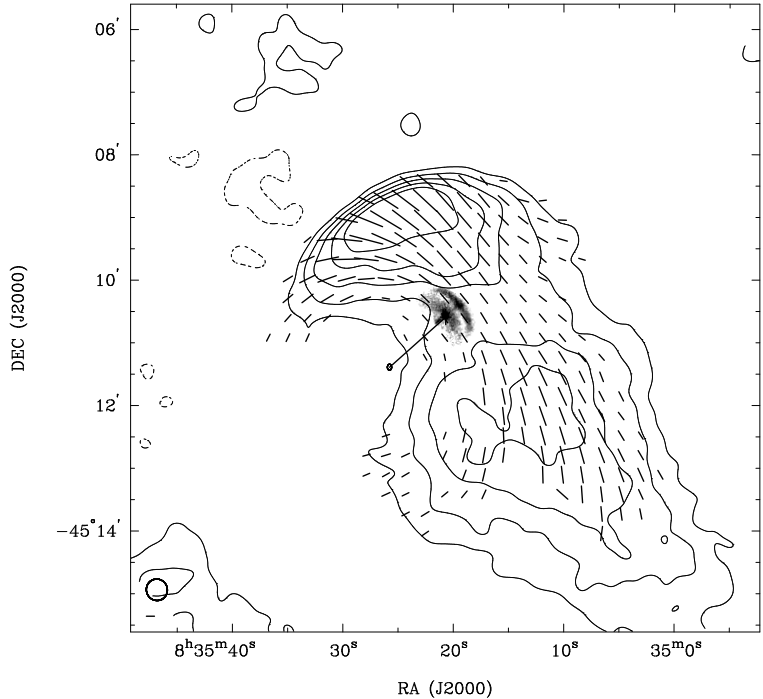}
\caption{{The Chandra observation of the Vela PWN (grey scale)
and, superimposed, the 5~GHz Radio contours (-1,1,2,3,4,5 mJy/beam
with $20^{\prime \prime}$ beam). The derotated magnetic field lines
are overlaid, with a 1~mJy bar at the bottom left, below the restoring
beam size. The proper motion vector shows the distance travelled in
1000 years, and ends with the three sigma error ellipse. The field
lines confirm that the radio PWN is created by the pulsar, and also
show the distortion from the motion of the pulsar.}}
\label{fig:vela}
\end{center}
\end{figure}

Why try to recover the polarisation of an observed radio source?
Primarily because it often yields unique information about the
source. Polarisation is a tracer of {\em asymmetry} in the source of
radiation, and the most common source of induced asymmetry is the
magnetic fields. `Seeing' the magnetic fields around a source of
interest provides huge insights into the source emission generation,
history and orientation in space. See for example figure
\ref{fig:vela} of the Pulsar Wind Nebula (PWN) around the Vela pulsar
(Dodson etal 2003) where the magnetic field, generated by the
particles spewing from the pulsar poles, curves around the pulsar and
is stretched back by the pulsar's motion through the Interstellar
medium (ISM).

%
Instrumental distortion of the polarisation is usually much greater
than the inherent signal itself. Polarisation calibration requires the
highest standards of general calibration, and it is this which makes
it difficult. I focus in this report only on the requirements to
correct the new types of telescope mounts. For a summary of the
reasons to observe polarisation and the methods see, for example, Ojha
(2001).

\section{Telescope Mounts}

Mount types are a combination of the focus position and the drive
type. In Radio Astronomy there are six more or less commonly used
focus positions, and three drive types. 

\subsection{Focus positions}

\begin{figure}[htb]
\begin{center}
\epsfig{file=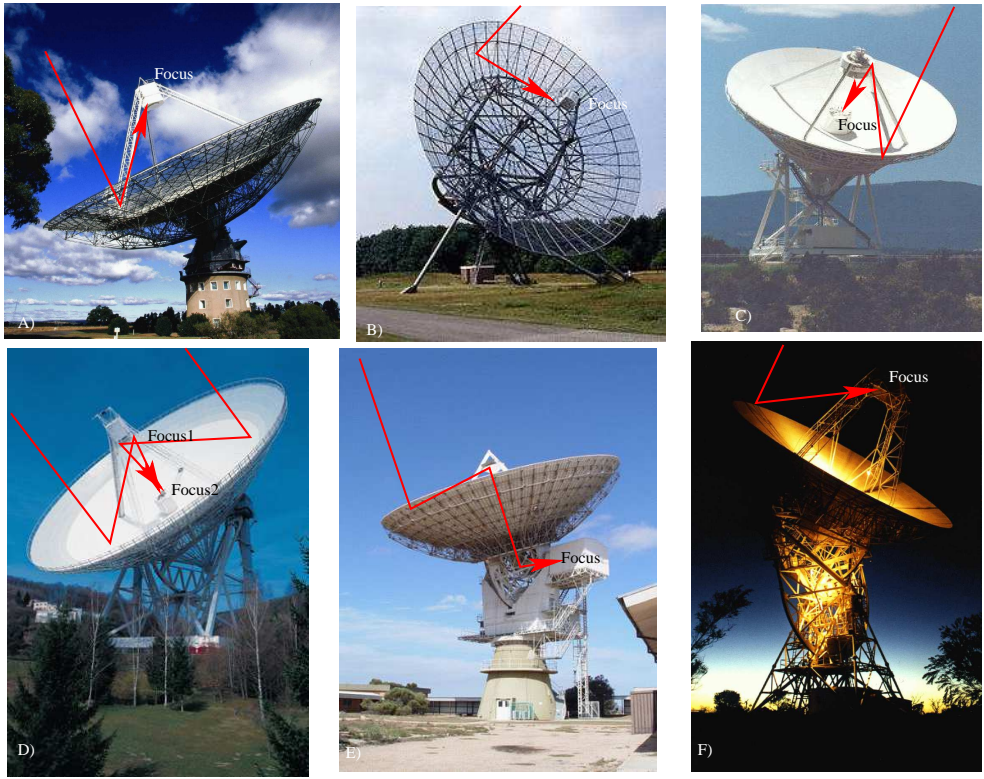, width=13cm}
\caption{{ A montage of different antennae displaying different
   focus positions and drive types. A) Parkes (NSW, Australia). Prime
   focus, Alt-Az drive, B) Westerborg (Holland). Prime focus, HA-Dec
   drive. C) Los Alamos (USA). Cassegrain focus, Alt-Az drive. D)
   Effelsberg (Germany) Either Prime or Gregorian focus, Alt-Az
   drive. E) Ceduna (SA. Australia). Folded Cassegrain focus, Alt-Az
   drive. F) Hobart (TAS Australia) Prime Focus, EW drive.}}
\label{fig:mounts_pictures}
\end{center}
\end{figure}

\begin{figure}[htb]
\begin{center}
\epsfig{file=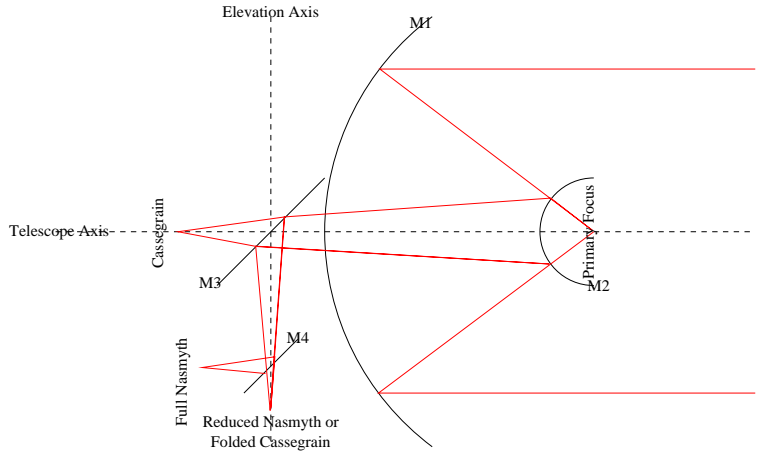,width=10cm}
\caption{Mount types}
\label{fig:mounts}
\end{center}
\end{figure}

Radio telescopes use a much more limited set of focus positions
compared to optical telescopes. Here we list them, along with examples
of the codes for telescopes which use them (in brackets). Images of these are
shown in figure \ref{fig:mounts_pictures} and a general schematic is
shown in figure \ref{fig:mounts}.

\begin{itemize}
\item Prime focus (PKS/MED). Focus at the site of the secondary mirror
  (sub-reflector).
\item Cassegrain focus (VLBA/most). Focus after the secondary
  (hyperboloid) mirror.
\item Gregorian focus (EFF). Prime Focus before the secondary
  (ellipsoid) mirror, secondary focus after the secondary mirror.
\item Folded Cassegrain focus (CED). Focus bolted to the elevation axis,
after the tertiary mirror. 
\item Reduced Nasmyth focus (JCMT). Focus on the elevation axis, but
bolted to the azimuth floor, after the tertiary mirror.
\item Full Nasmyth focus (PV/YEB40). Focus bolted to the azimuth axis
floor, after the forth mirror.
\end{itemize}
%

Mirrors after the forth, in a Nasmyth system, swap the mount type
between the equivalent of a reduced and full Nasmyth, so that any
system with an odd number of mirrors is a `reduced' and an even number
of mirrors is equivalent to a `full' Nasmyth system in our context. In
a similar fashion the expression for a folded Cassegrain and Prime
focus solutions are essentially the same, being either one or three
reflections from the on-sky view.
The difference between a Cassegrain and a Gregorian image is a
rotation (of 180$^o$) so these types are degenerate after the absolute
position angle calibration.
Furthermore at the correlator, the `on sky' left or right is combined
with the same to produce the parallel hand output. This is effectively
the same as adding another mirror to the optical chain for those
optics with an odd number of mirrors. For all expected cases,
therefore, we will be correlating either Cassegrain or Nasmyth foci,
combined with the drive type of the antenna. The other possibilities
were tested and included in the development code. They maybe of
use for single dish analysis, but it is not planned to include them in
standard AIPS.

\subsection{Drive configuration}

There are three drive configurations used in VLBI. This also effects the
rotation of the feeds as seen from the sky.

\begin{itemize}
\item HA-DEC (WST), for Hour Angle -- Declination. Also called
equatorial. It produces no change of feed angle as the source passes
across the sky. As HA-DEC mounts require an asymmetrical structural
design they are unsuited to the support of very heavy
structures. Their advantage is that motion is required in only
one axis to track an object as Earth rotates.
\item ALT-AZ (VLBA/most), for Altitude -- Azimuth. Also called
Az-El. It rotates the telescope pointing by the parallactic angle as
the source passes across the sky. It is the most common variety of
Radio telescope mount. It is symmetric so can support very large
antennae, but at the zenith small angular changes on the sky can
require very large angular changes in the azimuth angle. This effect
is call the `keyhole'.
\item EW (HOB), for East -- West. It rotates the telescope by the
co-parallactic angle as the source passes across the sky. An EW mount
places the focus very high, therefore it is prone to flexing and
poor pointing. Their advantage is that they can track across the sky at
high speeds and without gaps (i.e. it does not have a `keyhole' at the
zenith). They are primarily used for tracking Low Earth Orbit
satellites.
\end{itemize}

The parallactic angle, sometimes called the position angle, of an
object is the angle between the celestial pole (north or south
depending on location), the object and the zenith (the point in the
sky directly overhead). It is not to be confused for the same named
angle which is also called the convergence angle. This is the
difference in the angular direction to an object at a distance, from
two points of view, i.e. the parallax. The expression for the
parallactic angle as we use it is;

\[ 
\chi_p = {\rm arctan}[\frac{{\rm sin}(\Theta)\ {\rm cos}(l)}
                    {{\rm cos}(\delta){\rm sin}(l) -
		      {\rm cos}(\Theta){\rm cos}(l){\rm sin}(\delta)}]
\]

The co-parallactic angle is;

\[ 
\chi_c = {\rm arctan}[\frac{{\rm cos}(\Theta)}
                    {{\rm sin}(\delta){\rm sin}(\Theta)}]
\]

The Nasmyth angle is almost the same as the parallactic angle, but
with the relative rotation of the third mirror included;

\[ 
\chi_n = \chi_p \pm E
\]

where $\Theta$ is the hour angle, and $\delta$ the declination, of the
source. $l$ is the latitude of the telescope and $E$ is the
elevation. These angles are calculated in the code using the function
{\bf atan2} to allow the full range of angles to be returned.

\section{The Nasmyth Mount}

The feed in a Nasmyth system sits within the Azimuth cabin of an
Alt-Az telescope. Therefore the difference between the instrumental
polarisation angle of a Nasmyth mount and the parallactic angle is the
elevation angle. Consider the change in the image of the secondary
reflector projected onto the feed mount. At the horizon it has the
correct orientation, and is rotated as the telescope moves in
elevation. The sense of the rotation is positive (clockwise) or
negative (counter-clockwise) depending on whether the M3 (see figure
3) reflects to the right or left. The secondary reflector follows the
parallactic angle (assuming the drive type is an Alt-Az).
%
%
Each mirror adds a swap of polarisation from LHC to RHC. Therefore, in
principle, the number of reflections before the horn is very important
in a Nasmyth optical system, as each mirror also swaps the feed angle
polarisation vector on the sky.
However when the detected LHC or RHC polarisations are relabelled to
represent the on-sky value, an effective mirror is added to produce
the parallel or cross hand outputs. This means that the `reduced
Nasmyth' (and the `prime focus') case(s) should never be met in the
real world in VLBI. It maybe necessary if AIPS is used for single dish
reduction.
Nevertheless, the extension of the test version of AIPS to handle all
varieties of Nasmyth (three), Prime Focus and EW mount types was done
in parallel. The version which will be merged with the standard AIPS
will not include the extra code.

\section{The basic equations for the solution of polarisation}

There are numerous sources of information laying out the methods for
reducing polarisation observations. One of the best is EVN Memo 78
(Aaron, 1997), but also see Kemball (1999), Ojha (2001), Leppanen etal
(1995) and Hamaker (2000). All of these are for telescopes where the
feed angle is implicitly the parallactic angle (or possibly also the
equatorial angle). If one replaces all occurrences of the term
`parallactic angle' with the more generic `feed angle' these are all
correct. From these references we repeat only the derivation of the
D-term equations.

For an ideal case (i.e. unitary gain, no cross talk and no on-sky feed
rotation) the observables are:

\[
R_1 R_2 = (I_{12} + V_{12}) 
\]\[
L_1 L_2 = (I_{12} - V_{12}) 
\]\[
R_1 L_2 = (Q_{12} + i U_{12}) = P_{12} 
\]\[
L_1 R_2 = (Q_{12} - i U_{12}) = P^*_{12} 
\]

Where R and L are Right and Left handed polarisations, I, Q, U and V
are the stokes parameters, and P is the linearly Polarised flux.

But when we have gain terms ($g$), cross talk ($R_{obs}=R + DL$), and feed
rotation ($\chi$) this becomes, for example;

\begin{displaymath}
\begin{array}{lr} R_1L_2 = <E_{R_1}E_{L_2}^*>  =
                      & g_{R_1} g^*_{L_2}[ P_{12}
                        e^{-i(\chi_1+\chi_2)} + \\
                      & D_{R_1} D^*_{L_2} P^*_{12}
                        e^{i(\chi_1+\chi_2)} + \\
                      & D_{R_1} L_1 L_2 e^{i(\chi_1-\chi_2)} + \\
                      & D^*_{L_2} R_1 R_2 e^{-i(\chi_1-\chi_2)}]
\end{array}
\end{displaymath}

Where $g$ is the complex gain, $\chi$ the phase from feed angle, and
$D$ the D-terms. However, assuming $V=0$ (i.e. zero circular
polarisation), and that small terms (e.g. $D^2$) tend to zero, we
can simplify to:

\[
R_1 R_2 = g_{R_1} g_{R_2}^* I e^{-i (\chi_1 - \chi_2)}
\]\[
L_1 L_2 = g_{L_1} g_{L_2}^* I e^{+i (\chi_1 - \chi_2)}
\]\[
R_1 L_2 = g_{R_1} g_{L_2}^* [P e^{-i (\chi_1 + \chi_2)} 
+ D^*_{L_2} RR  e^{-i (\chi_1 - \chi_2)}
+ D_{R_1} LL  e^{i (\chi_1 - \chi_2)} ] 
\]\[
L_1 R_2 = g_{L_1} g_{R_2}^* [P^* e^{i (\chi_1 + \chi_2)} 
+ D^*_{R_2} LL  e^{+i (\chi_1 - \chi_2)}
+ D_{L_1} RR  e^{-i (\chi_1 - \chi_2)} ]
\]

Assuming that $P$ is a constant fraction of $I$ ($P=pI$), and $RR=LL=I$
(as $V=0$), and exchanging $g{^\prime}_{R_m}$ for $g_{R_m} e^{-i
\chi_m}$ and $g{^\prime}_{L_m}$ for $g_{L_m} e^{+i \chi_m}$ we get;

\[
R_1 R_2 = g{^\prime}_{R_1} g{^\prime}_{R_2}^* I ; L_1 L_2 = g{^\prime}_{L_1} g{^\prime}_{L_2}^* I
\]
\[
R_1 L_2 = g{^\prime}_{R_1} g{^\prime}_{L_2}^* I (p + D^*_{L_2} e^{2 i \chi_2}
+ D_{R_1} e^{2 i \chi_1} ]
\]
\[
L_1 R_2 = g{^\prime}_{L_1} g{^\prime}_{2_2}^* I (p^* + D^*_{R_2} e^{-2 i \chi_2}
+ D_{L_1} e^{-2 i \chi_1} ]
\]

The polarisation calculation is based on the fit of the cross to
parallel flux ratios: e.g. $RL/RR$. These are: 

\[RL/RR = \frac{g{^\prime}_{L_2} g{^\prime}_{R_1} I}
{g{^\prime}_{R_2} g{^\prime}_{R_1} I} 
(p + D^*_{L_2} e^{2 i \chi_2} + D_{R_1} e^{2 i \chi_1} )
\]

\[ = p e^{-2 i \chi_2} + D^*_{L_2} +  D_{R_1} e^{2 i (\chi_1-\chi_2)}
\]

similarly,  

\[LR/RR = p^* e^{2 i \chi_1} + D_{L_1} +  D^*_{R_2} e^{2 i (\chi_1-\chi_2)}
\]

\[RL/LL = p e^{-2 i \chi_1} + D_{R_1} +  D^*_{L_1} e^{-2 i (\chi_1-\chi_2)}
\]

\[LR/LL = p^* e^{2 i \chi_2} + D^*_{R_2} +  D_{L_1} e^{-2 i (\chi_1-\chi_2)}
\]

A final enhancement is that P, the polarised emission, can be modeled
as polarised fractions of {\em subsections} of the total
emission. That is;

\[P=\Sigma_n p_n I_n\]

This is the formalisation used in the AIPS task LPCAL, and is why
LPCAL is to be preferred for VLBI polarisation calibration.

\section{Formulation in terms of Jones matrices}

Jones Matices are a very useful way to express the optical chain. The
usual formalisation, and the one used here is for the X and Y
(Elevation and Azimuth) axes. 

A mirror (reflecting from left to right) is:

\begin{displaymath}
M = \left( \begin{array}{rr}1 & 0 \\0 & -1 
\end{array} \right)
\end{displaymath}

A rotation of $\theta$ (where $\theta$ is the elevation in the Nasmyth
case) is:

\begin{displaymath}
R = \left( \begin{array}{rr}cos(\theta) & sin(\theta) \\
-sin(\theta) & cos(\theta)
\end{array} \right)
\end{displaymath}

The complex D-terms
are:

\typeout{Check the D terms}
\begin{displaymath}
D = \left( \begin{array}{cc}1&D_{x}\\D_{y}&1
\end{array} \right)
\end{displaymath}

For the case of a Full Nasmyth antenna the Jones matrices include the
Cassegrain terms from the `structure' (following the parallactic
angle) $D_C$ and the Nasmyth terms from the `feeds', $D_N$. The
resultant optical chain is equivalent to: 

\[  \equiv (M_n \cdots M_5) \cdot D_N \cdot M_4 \cdot R(t) M_3 \cdot
 [ D_C \cdot M_2 \cdot M_1 ] \] 

From this formulation it is clear that the $D_N$ and $D_C$ can not be
combined, as they are either side of the time varying rotation matrix.
Furthermore is obvious that if there is a final number of odd mirrors
the solution is equivalent to a reduced Nasmyth system, and the full
Nasmyth applies to the case with an even number of mirrors. But recall
that exchanging the labels of LHC and RHC to reflect the on-sky
polarisation is the equivalent to yet another mirror.
Finally that the D-term solutions will swap (to their negated
conjugate) in each mirror, if the calibrator is unpolarised (or
assumed to be unpolarised). The expression in square brackets is that
for a Cassegrain system.

\section{Contributions to the D-terms}

D-terms arise from the cross talk of the nominally right hand circular
feed to the nominally left hand circular feed (or X and Y linear
feeds, which are not relevant to the discussion here). Three dominant
mechanisms for cross talk exist: i) {\underline {Elliptical feeds}}:
In practice linear probes are used to detect the incoming wave,
therefore a quarter wave plate is placed in front of these feeds to
convert the incoming circular polarisation to linear polarisation. A
quarter wave plate is only perfect at a single frequency, and even
then only if perfectly constructed. Therefore the conversion will
always produce a certain amount of ellipticity which is tied directly
to the receiver; ii) {\underline {Impedance-matching}}: The feeds will
not be perfectly impedance-matched to the incoming system, so there
will be a certain amount of power re-emitted from each probe. This
power can then be reflected back into the receiver, with the circular
polarisation swapped, and this reflection can be from anywhere in the
optical system. This is not normally an issue, as long as they are
constant in time, as these D-terms are degenerate with those of the
receiver. In a Nasmyth antenna this is not true for reflections
returning from the M2 mirror as there is a time varying rotation at
this point; iii) {\underline {Pointing offsets}}: Even a perfect
system has intrinsic D-terms away from the symmetry axis (because of
the loss of symmetric cancellation of terms). These are inextricably
tied to the Alt-Az mount through the sub-reflector (M2), and in
principle could change with time. These, if significant, will have to
be solved for using the Az-Alt mount type, see Brisken (2003) for the
eVLA analysis. However modeling at Yebes (F. Tercero, personal comm.),
particularly for the SX receiver package, has suggested that the level
of cross talk will be less than 0.1\%. These levels, if correct, will
be undetectable.
Time varying D-terms would require unpolarised calibrators as is
impossible to get good parallactic angle coverage in short time
spans. The Left-Right `beam squint' also is a contributor to off axis
polarisation for the VLA. In the usual VLBI case, because the fields of view
is so much smaller, it does not contribute. The Left-Right gains are
solved for at the pointing centre and no imaging is attempted
significantly far from this point. However, for wide field VLBI
imaging, it will become an issue. In this case amplitude calibration
would need to be performed at each imaging centre.

The need for double D-terms are conceivable, therefore we have
developed test code and a method to allow combined Nasmyth and
Cassegrain D-term solutions. It is not clear that the secondary terms
will be necessary in any `real-world' example.

\section{Double D-terms}

Returning to the expressions for the cross hands with D-terms for two
kinds of rotation we find: 

\begin{displaymath}
\begin{array}{lllll} 
R_1 L_2 = & g_{R_1} g_{L_2}^* [P e^{-i (\chi_1 + \chi_2)} &+\\
& D^{*}_{L_2} RR  e^{-i (\chi_1 - \chi_2)} &+ 
& D^{\prime *}_{L_2} RR  e^{-i (\chi_1 - \Theta_2)} &+ \\
& D_{R_1} LL  e^{i (\chi_1 - \chi_2)} &+ 
& D^\prime_{R_1} LL  e^{i (\Theta_1 - \chi_2)}& ~] 
\end{array}
\end{displaymath}

\begin{displaymath}
\begin{array}{lllll} 
L_1 R_2 = & g_{L_1} g_{R_2}^* [P^* e^{i (\chi_1 + \chi_2)} &+\\
& D^*_{R_2} LL  e^{+i (\chi_1 - \chi_2)} &+
& D^{\prime *}_{R_2} LL  e^{+i (\chi_1 - \Theta_2)} &+\\
& D_{L_1} RR  e^{-i (\chi_1 - \chi_2)} &+
& D^\prime_{L_1} RR  e^{-i (\Theta_1 - \chi_2)} &~]
\end{array}
\end{displaymath}

Where $\chi$ is the feed angle, and $\Theta$ is the extra (parallactic
angle) term, and $D^{\prime}$ is the $D$ term associated with this extra term.

This, after the feed angle correction, reduces to: 

\[
R_1 L_2 = g{^\prime}_{R_1} g{^\prime}_{L_2}^* I (p +
 D^*_{L_2} e^{2 i \chi_2} + D_{R_1} e^{2 i \chi_1} +
 D^{\prime *}_{L_2} e^{i (\Theta_2 + \chi_2)} +
 D^\prime_{R_1} e^{i (\Theta_1 + \chi_1)} ]
\]
\[
L_1 R_2 = g{^\prime}_{L_1} g{^\prime}_{2_2}^* I (p^* + 
D^*_{R_2} e^{-2 i \chi_2} +
D_{L_1} e^{-2 i \chi_1} +
D^{\prime *}_{R_2} e^{-i (\Theta_2+\chi_2)} +
D^\prime_{L_1} e^{-i (\Theta_1+\chi_1)} ]
\]

The easiest way to investigate this effect was to allow the fitting of dual
terms -- but only to apply one. A second pass of LPCAL would then
allow the application of the other term, with the other feed
angle. Therefore, in the code (LPCAL\_EXT) only one ``double termed''
antenna.

\section{AIPS code changes}

The fundamental alteration was to add support for the Nasmyth optics
(and EW-mount optics). The core subroutine for the calculation of
parallactic angles was extended to return either, the parallactic plus
or minus the elevation angles, or co-parallactic angles (i.e. the
correct formalisation for these mount types). In addition test code
was written for reduced Nasmyth and Prime focus mount types. The mount
types, and their numeric values, are listed in table 1.

\begin{footnotesize}
\begin{table}[htb]
\begin{center}
\begin{tabular}{lcrr}
&&\multicolumn{2}{c}{Mount Number}\\
Focus and Drive&Label&Test& Final\\
\hline
Cassegrain and Alt-Az& ALAZ &0&0\\
Any and Equatorial & EQUA &1&1\\
Any and Orbiting& ORBI &2&2\\
\multicolumn{3}{c}{New mount types}\\
Prime focus and EW  & EW-\,- &$\pm$3&3\\
Right hand Naysmyth and Alt-Az& NS-R &4&4\\
Left hand Naysmyth and Alt-Az& NS-L &-4&5\\
\multicolumn{3}{c}{Extra mount types}\\
Reduced Nasmyth and Alt-Az& R-NS &$\pm$5&-\\
Prime focus (or Folded Cassegrain) and Alt-Az& PRIM &6&-\\
\hline
\end{tabular}
\caption{Mount Types}
\end{center}
\end{table}
\end{footnotesize}

The extra mount types will not be carried forward for inclusion in to
AIPS, as they are only for testing. 

All calls to this code have been checked, as have all calls to the
mount type via GETAN, all mentions of ``parallactic'' in the source
code, and all mentions of any of the common names for mount type
(MNTYP, MNTSTA, MNTEL and IDTMNT). All issues associated with those
have been fixed. A full list of the differences can be found in
appendix \ref{app:code}. 

The greatest problem was that not all derivations of the parallactic
angle were done with a single common subroutine. This was fixed for
most cases, but in one case it has not been standardised, i.e. VPLOT
(for which one can use UVPLT). In VPLOT the position angle is only
used as an alternative plotting axis, and therefore has no impact on
the data reduction.

A point of concern is that these corrections are for a `moving
target'. The code written was for the 31DEC05 version of AIPS
downloaded in February 2005. There have been considerable changes
since that date. The latest version was for the January version of
31DEC07.
It is highly desirable that the new code is merged with the
distribution tree.

\section{Data visualisation}

 The first stage of calibration is to correct the relative phase
 rotation between the left hand and the right hand receivers. The success of
 this stage can be easily demonstrated with the task VPLOT by plotting the
 phases difference between the two polarisations (usually against time).
 The phase of the ratio of the parallel hands will be, before feed
 angle calibration;
\[
\frac{R_1 R_2}{L_1 L_2} = \frac{g_{R_1} g_{R_2}^*}{g_{L_1} g_{L_2}^*}
e^{-i (\chi_1 - \chi_2)}
\]
and after;
\[
\frac{R_1 R_2}{L_1 L_2} = \frac{g{^\prime}_{R_1} g{^\prime}_{R_2}^*}
                               {g{^\prime}_{L_1} g{^\prime}_{L_2}^*}
\]
Therefore the phases between the two polarisations, post feed angle
calibration, will have a constant phase. See figure
\ref{fig:br46_rrll}. Post phase calibration, of course, this phase
will be zero.
 This step is in theory independent of any other calibration but in
 practice, as the data is averaged then compared, the delays need to
 be solved for and the VPLOT averaging has to be less than the
 (post-calibration) coherence time. That is, either short averaging
 intervals must be used (with the risk that the phase difference will
 be undetectable) or at least some phase calibration must have been
 performed. It is important that, in this case, the phases
 applied to each polarisation are not independent. Otherwise the
 effect is masked as the phases are absorbed into the
 calibration. Examples of these RR/LL plots are to be found in figures
 \ref{fig:br46_rrll} and
 \ref{fig:c051a_rrll}.

 Conventionally the quality of the polarisation solutions was judged
 from plotting RL/RR (or LR or against LL) on the real/imaginary plane
 with and without polarisation calibration. Without calibration the
 values will fall on a circle around an (offset and potentially
 moving) origin, as a function of feed angle difference, following:

\[ RL/RR= p e^{-2 i \chi_2} + D^*_{L_2} +  D_{R_1} e^{2 i (\chi_1-\chi_2)}
\]

\begin{figure}[htb]
\begin{center}
\epsfig{file=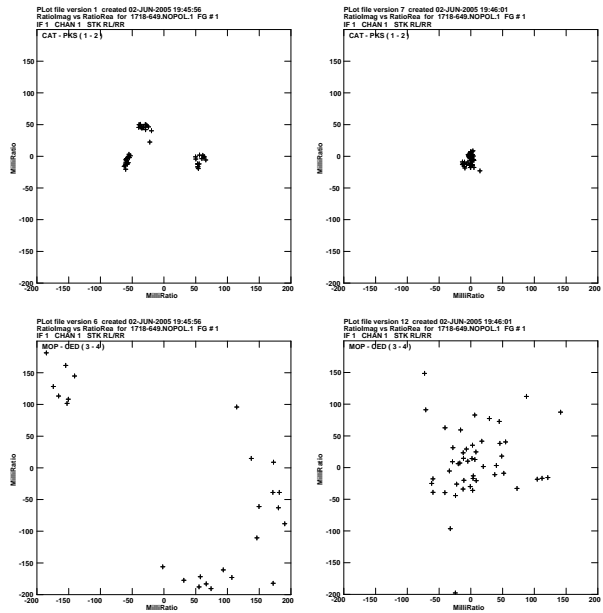,angle=0,width=8cm}
\caption{Pre- and Post- polarisation solution from traditional VPLOT
  display for the four antennae in V148A.}
\label{fig:vis_trad}
\end{center}
\end{figure}

\begin{figure}[htb]
\begin{center}
\epsfig{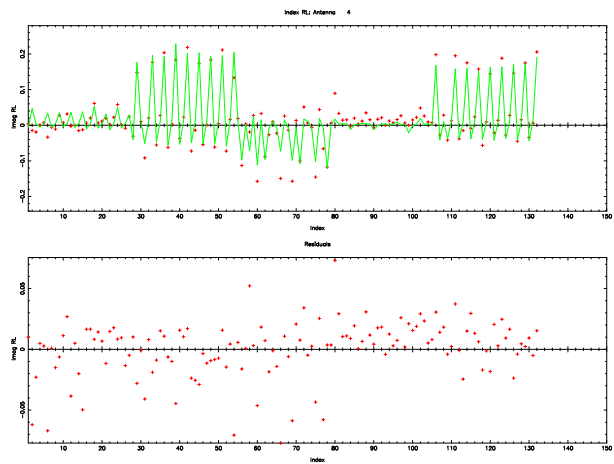}
\caption{Solutions for Ceduna in experiment V148A displayed via the
new interactive visualisation subroutines. The shown visualisation
mode is showing imaginary RL flux (in Jy) against baseline and time
(index). The upper window shows the data (in red points) and the model
(with a green line) and the lower shows the residuals. For each
integration period the data for each baseline involving the selected
antenna (number 4, Ceduna) is plotted.}
\label{fig:vis_new}
\end{center}
\end{figure}

\begin{figure}[htb]
\begin{center}
\epsfig{file=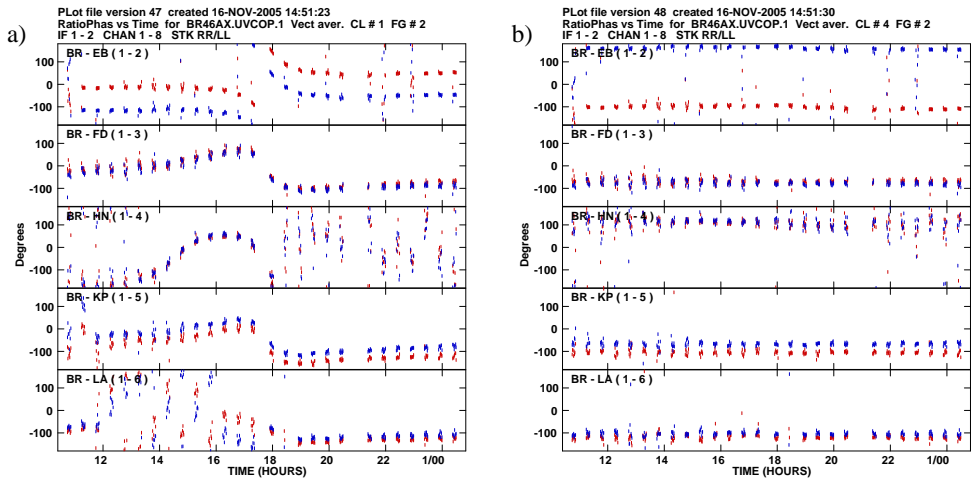,width=13cm}
\caption{Plots demonstration the successful correction of the
  differential phase rotation of the two parallel hands for experiment
  BR046. a) is pre-calibration and b) is post calibration.}
\label{fig:br46_rrll}
\end{center}
\end{figure}

 I.e. if the source is unpolarised ($p=0$), the centre is $D^*_{L_2}$
 and the radius of the circle is $D_{R_1}$. Examples of such plots are
 to be found in figure \ref{fig:vis_trad} etc. This is a useful
 diagnostic tool, but it can be improved on. I added code to LPCAL
 which allows the interactive plotting of the data, with and without
 the model fitted removed, against many other data axes, see figure
 \ref{fig:vis_new}.

 The extended data visualisation is done via a PGPLOT X-window
 interface added into LPCAL. Therefore LPCAL needs to be compiled with
 extra options. This is not planned to become part of classic
 AIPS\footnote{The PGPLOT copyright means that this probably cannot be
 part of a general AIPS release. PLPOT is an alternative library which
 would avoid this problem.} but can be requested. The data is plotted
 (as coloured points) in several forms with the model overlaid (as
 coloured lines). The residuals are displayed in a subplot. See Figure
 \ref{fig:vis_new}. Plots of RL or LR data, can be on the complex
 plane or against data index or feed angle. Either the absolute, real
 or imaginary data can be plotted. Individual antennas can be
 selected, or overlaid.

\subsection{Commands}

\begin{itemize}
\item Data Plane: Complex (c), feed Angle (a) or array Index (i)
\item Data product: RightLeft (r) or LeftRight (l)
\item Data Value: Absolute (1), Real (2) or Imaginary (3) 
\item quit (q)
\end{itemize}
Extras: 
\begin{itemize}
\item Site selection (s followed by number).\\
When greater than 9 use + and number (i.e. s+1 for 11). (Tip: s++
increments antenna number by 1)
\item Model on or off (m)
\item Differential value (d) (i.e. $\chi_i-\chi_j$ for feed angle, or
baseline index rather than total index)
\item Zoom: Top corner (t) and Bottom corner (b). Reset with new data
\item Plot: a colour postscript (pgplot.ps) file is generated (p) of the
current viewing plane.
\end{itemize}

\subsection{Comments}

The Index data plane, which I personally find most useful, will
usually represent the data in the ``TB'' sort order. This leads to the
raggedy appearance, as each baseline is looped over for every time
index. Altering the order to ``BT'' would remove this effect, but
LPCAL does not work on BT ordered data. No solution for this has been
identified. 

The Zoom needs to be selected from the lower plot. The default limits
come from the maximum and minimum of the entire dataset. With a SOLINT
of 1 minute this produces a great scatter. 

\clearpage

\section{Demonstration of a conventional VLBI data analysis}

\begin{figure}[htb]
\begin{center}
\epsfig{file=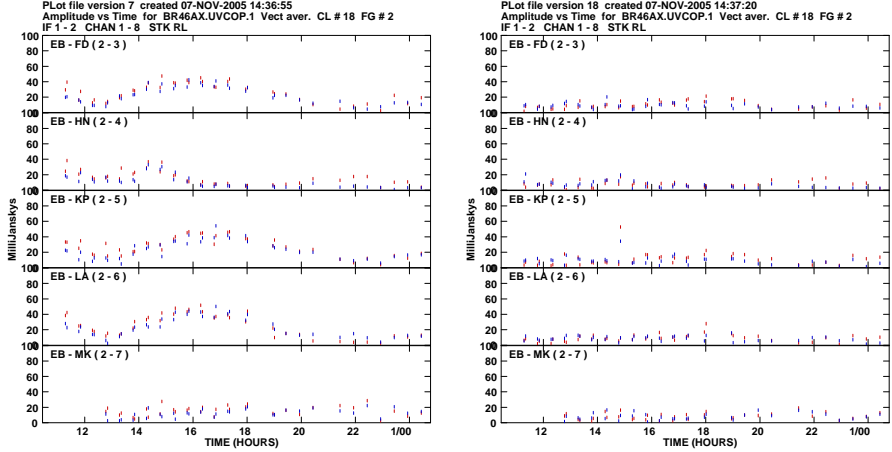,width=12cm}
\caption{Selected plots from the task VPLOT showing the successful
 polarisation calibration of experiment BR046. The pre polarisation
 calibration is shown on the left for a number of telescopes to
 Effelsberg. Post calibration is on the right. Note that the magnitude
 of the flux in the cross hands falls, and becomes constant with time.}
\label{fig:vis_46}
\end{center}
\end{figure}

It is important to confirm that our changes have not broken the AIPS
system. I have use the VLBA data set BR046. The Figures
\ref{fig:vis_46} show the solutions.
We find the same solutions as those derived on a vanilla AIPS
installation, plus on the target and calibration sources (to within
errors). Furthermore we find that the solutions are robust when
derived from the other source, and with the assumption of an
unpolarised point source. We conclude that the changes are non-toxic.

Furthermore we analysed the EVN dataset N05L1, a network monitoring
experiment. This included OQ208 which is unpolarised, and therefore
could be used as a calibrator despite the short length (and therefore
poor parallactic angle coverage) of the experiment. In this dataset of
thirteen antennas there were three antenna with prime foci and two
with equatorial mounts. The best solutions were found following the
conventional analysis (as would be expected), but we also tested the
reduction with the prime focus mount type.

\begin{figure}[htb]
\begin{center}
\epsfig{file=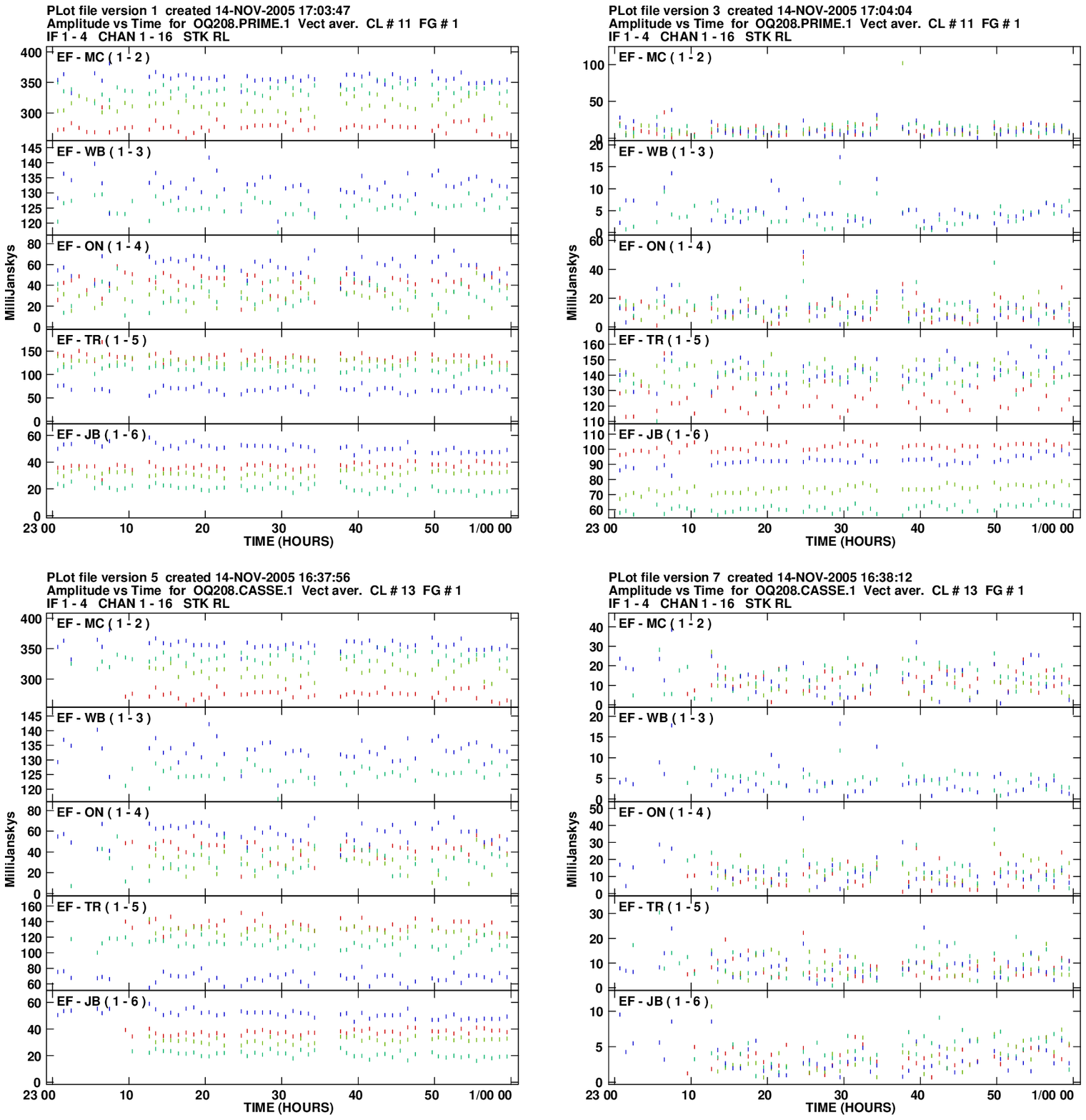,width=11cm}
\caption{Selected plots from the task VPLOT showing the successful
 polarisation calibration of experiment NL051. The pre polarisation
 calibration is shown on the left for a number of telescopes to
 Effelsberg. Post calibration is on the right. The upper plots are for
 analysis with the PRIME focus used for Effelsberg, Medicina and
 Jodrell Bank, the lower used the nominal mount types. The solutions
 based on the nominal types are significantly better.}
\label{fig:vis_51}
\end{center}
\end{figure}

\section{Investigations of Prime Focus, Folded Cassegrain and
EW-mount VLBI data analysis}

This required the extra step of installing ATLOD to AIPS. It is
required for all tasks involving the Australian CSIRO ``Radio Physics
Fits'' (RPF) format. I used the experiment LBA data set V148. This
experiment of a 6.7 GHz polarisation experiment with Parkes (Prime
focus), ATCA and Mopra (Cassegrain), Ceduna (Folded Cassegrain) and
Hobart (E-W mount). It was in two session (A1+A2 and B) and included
seven scans each of two polarisation calibrators (1610-771 and
1718-649) which allowed the demonstration that:

\begin{itemize} 
\item The additional mount types did not effect the conventional
mounts. 
\item That the zeroth order Nasmyth (the folded Cassegrain) and the
Prime focus have D-terms that are negated and conjugated if they are
treated as a conventional Cassegrain system.
\item That the number of antennae needs to be greater than three to
also solve the polarised fraction of the source \footnote{The number
of D-terms are twice the number of antennae. The number of cross-hands
used are twice the number of baselines, the number of source
polarisations are the number of regions with polarised emission
assumed in the image (number of model components, or the number of
clustered clean components). If the source is assumed to have a single
polarisation fraction the number of unknowns is greater than the number
of observations for three antenna. For four there can be up to four
regions of polarisation, for more this is rarely an issue.}.
\item That the antennae Mopra, Parkes and ATCA have very similar
parallactic angles, and therefore the solutions with just these maybe
degenerate. The addition of (sufficient) additional antennae breaks
this degeneracy.
\end{itemize} 


%


Dataset V148A raised some interesting questions as stable solutions
are only found when Parkes is treated as a prime focus and Ceduna as
(the equivalent) folded Cassegrain. This contradicts the results from
the EVN (where Effelsberg is a prime focus, but the most stable
solutions were found when it was treated as a Folded Cassegrain
(i.e. the sign reversed) feed). No explanation was found for this, and
as reasonable solutions were found for the conventional analysis, we
decided this was an unexplained distraction which maybe related to the
limited number of antennae.

\begin{figure}
\begin{center}
\epsfig{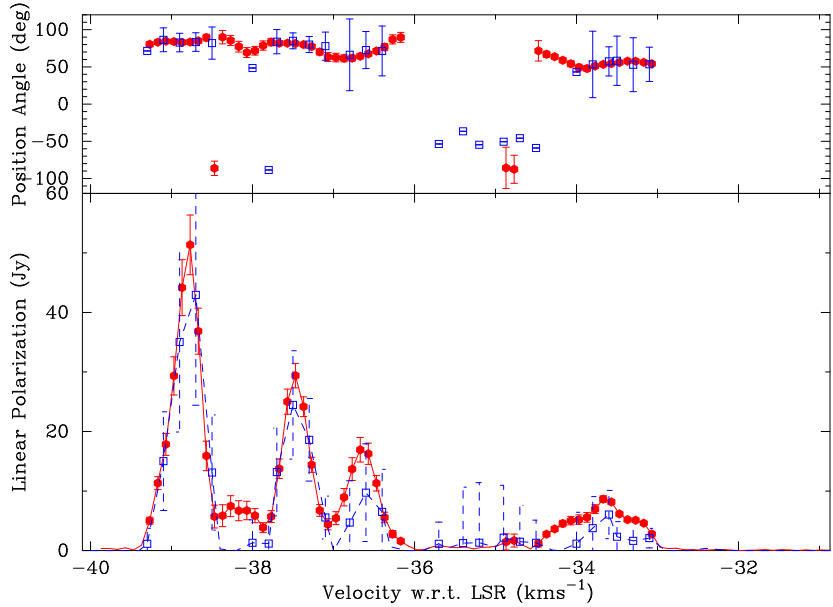}
\caption{Polarisation angle and fraction for G339-1.26, as observed by
  the LBA (this paper, blue open squares) and the ATCA (Ellingsen,
  priv. comm., red closed circles). The spectra is scalar summed
  across the image (Stokes I,Q and U), and shows good agreement
  between the VLBI and the connected array results. The errors are the
  absolute errors based on the confidence in the polarisation
  calibration (2\% and 0.4\% respectively), not the relative
  errors. Where errors are not shown they could not be
  calculated. From Dodson, 2007}
\label{fig:pa}
\end{center}
\end{figure}

V148B behaved in a more sensible manor, and we note that this dataset
has one additional antenna (Hobart) which might be providing enough to
break the degeneracy. This underlines the importance of the EW mount
type code; without it the (usual) LBA has insufficient differential
parallactic angle coverage. Figure \ref{fig:pa} shows the comparison
of the position angles found from the V148B experiment and those from
an ATCA observation of the same source. They agree very well, with the
added benefit of the VLBI observations resolving separate components
and allowing better polarised flux recovery. The polarised flux found
in the VLBI observations between -35 and -36\,kms$^{-1}$ have, in fact,
better recovery of polarised flux then the ATCA which does not show up
in the {\bf imspec} generated plots. These velocities are the overlap
region between the Eastern and Western clusters, so the lower
resolution of the ATCA is blending these and thus not detecting the
polarisation.

\section{D-terms for experiment V182A}
\label{app:v182}

This experiment (PI Dr Dodson) was performed on the LBA at 4.8 GHz
with 64 channels of about 0.25~MHz each. The target source (J0743-67)
is an AGN with an interesting double structure. The calibrator was
J0637-752, also an interesting source. 
It is the first experiment to produce solutions for the polarisation
on the LBA, but was not designed as a polarisation experiment,
therefore it does not include a absolute polarisation calibrator and
all position angles are arbitrary. It included Hartebeesthoek so had
more baselines to solve for amplitude and D-terms. Reduction followed
the standard routes, apart from changing the mount types to EW mount
for Hobart and equatorial mount for Hartebeesthoek.

\begin{figure}[htb]
\begin{center}
\epsfig{file=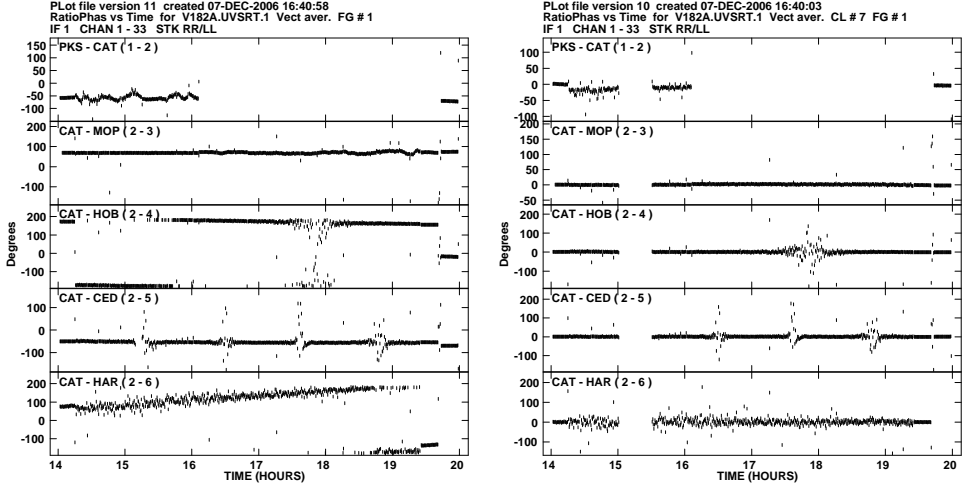, width=13cm}
\caption{The RR/LL phase difference in LBA experiment V182A before and
after feed angle correction types. The mount types are parallactic and
equatorial and co-parallactic. The phase between the two polarisations
before (left) and after removal of the feed rotation in V182A, for all
antennae to the CAT. The rotation of the feeds introduces a variable
phase between the two hands in the data. The effect of the near
identical feed angles for the NSW antenna is indicated by the near
constant phases between them. The phases are flattened (and zeroed for
the calibrator at 14UT) for all antennae in the corrected data, which
is shown on the right.}
\label{fig:hob_test}
\end{center}
\end{figure}

\begin{figure}
\begin{center}
\epsfig{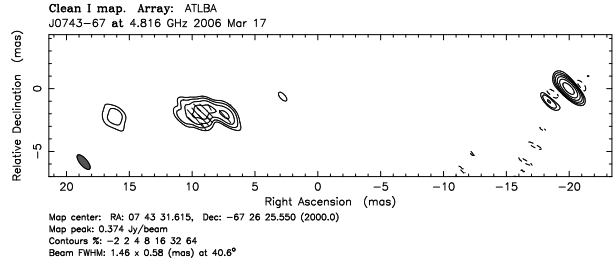}
\caption{The image of J0743-67 from experiment V182A. The polarisation
  vectors are shown overlaid. Absolute polarisation angles can not be
  derived so the direction is arbitrary. The core (to the West) is
  unpolarised ($\le 1\%$) and the jet (to the East) is smoothly
  polarised with a polarised fraction of approximately 16\%.}
\end{center}
\end{figure}



\section{Demonstration of a Full Nasmyth optics VLBI data analysis}

\begin{figure}[htb]
\begin{center}
\epsfig{file=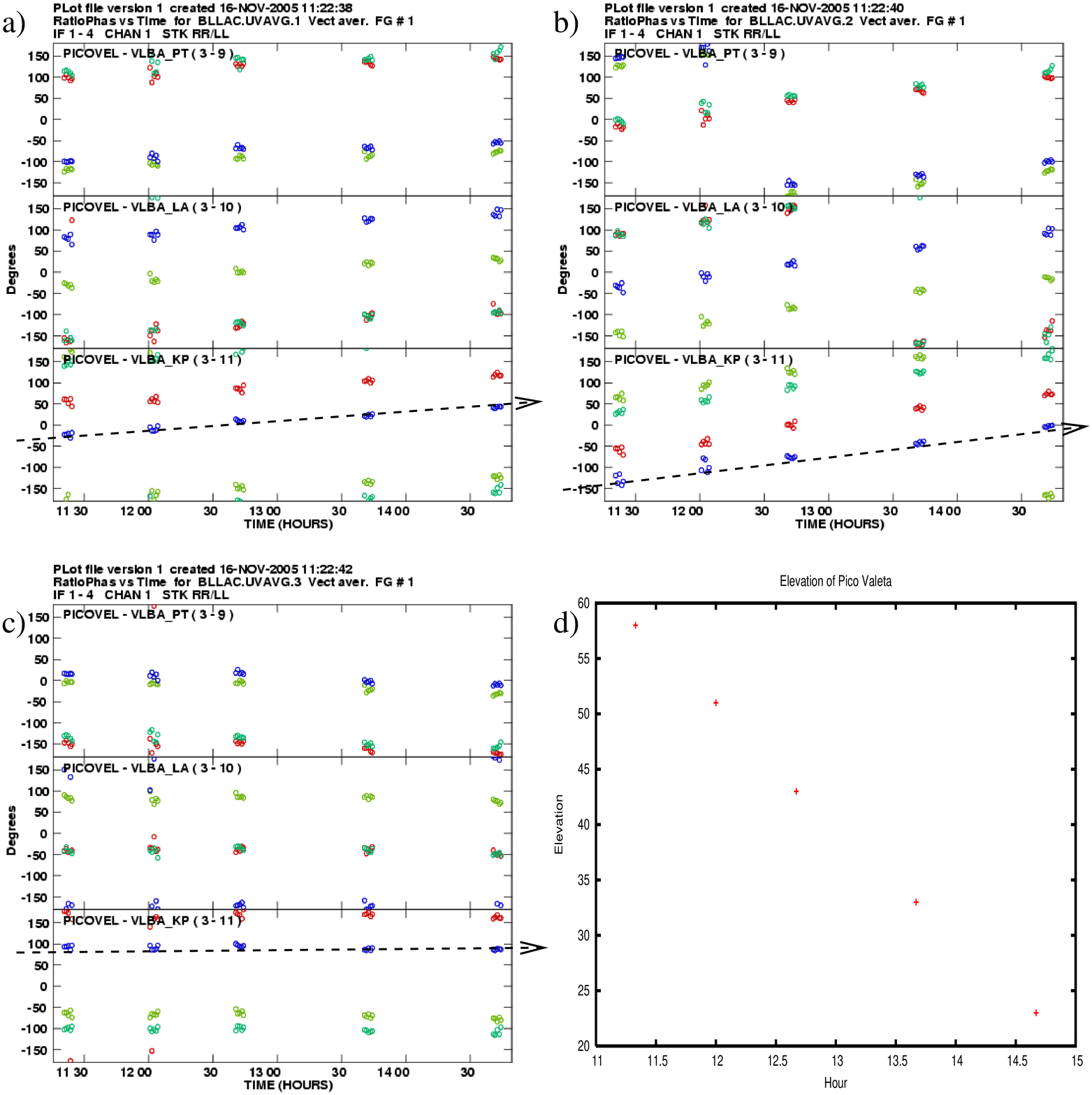,width=12cm}
\caption{The RR/LL phase difference in GMVA experiment C051A for three
parallactic angle correction types: a) Cassegrain, b) Nasmyth right handed
and c) Nasmyth left handed. d) Shows the elevation angle for these
observation times. The dashed line proves a guide to the eye of phase
changes with time. It is obvious that case c) is correct and has
flattened the phase response.}
\label{fig:c051a_rrll}
\end{center}
\end{figure}

During the project the only suitable data from the Pico Veleta antenna
that was obtained was without any amplitude calibration or flagging
tables. Nevertheless I was able to demonstrate the successful
correction of the feed angle terms. Three types of correction were
applied. Those for Cassegrain feed rotation (mount type 0, or ALAZ)
those for the Right-handed Nasmyth feed rotation (mount type 4) and
those for left-handed Nasmyth feed rotation (mount type 5). Figure
\ref{fig:c051a_rrll} plots the phase difference between left and right
hand polarisation for each of these cases, plus the elevation of Pico
Veleta at the times of observation. The slope without Nasymth
correction is positive, and the opposite of the trend of the
elevation, with the incorrect Nasmyth correction the slope is greater,
and with the correct terms the phase difference is flat with time.
Further analysis of this experiment is impossible, as amplitude
calibration cannot be performed and, as stated, that is an essential
precondition for good polarisation calibration. This, however, is in
itself a complete demonstration of the application of the Nasmyth feed
angle expression. Combined with the successful application of other
feed angle expressions the calibration pipeline is demonstrated to be
complete.

\clearpage
\begin{footnotesize}

\end{footnotesize}

\pagebreak

\appendix

\section{Recommendations for calibration of polarisation observations}

Many references have covered this, but here I wish to clearly layout
the important steps in AIPS and the consequences of the tasks.

\begin{itemize}
\item {\bf TABED} options: INE 'AN'; OPTY 'repl'; APARM 5 0 0 4 4 {\bf
  3} and KEYV {\bf 5} 0\\ Replaces the MNTSTA type of antenna 3 with
  value 5 (for a left handed Nasymth, such as Pico Veleta).\\
  For Hobart and the LBA this needs to be type 3, furthermore one
  needs to change for all antenna: columns 8 and 11 to zero (APARM
  8/11 0 0 2 0;KEYV 0), column 7 to 'R' and 10 to 'L' (APARM 7/10 0 0
  3 0;KEYST 'R'/'L'). Recall also that Harts is Equatorial (mntyp 1).
\item {\bf CLCOR} options: CLCORP 1 0 and OPC 'pang'\\
Calculates the phase correction for the listed mount types.
\item {\bf FRING} options: CALS 'prime\_cal'; APARM(3)=0 and APARM(5)=0\\
Finds the independent delays and combined rates and phases for the
calibrator. Now the RR/LL phase will be constant.  
\item {\bf CALIB} options: CALS 'prime\_cal'; SOLMODE 'P'; APARM(3)=0 and
APARM(5)=0\\ 
Finds the independent rates and phases for the calibrator. Now the
RR, LL and RR/LL phases will be zero. One may wish to use only one
scan on the prime calibrator. 
\item {\bf VLBACPOL} this procedure finds the delays between left and
  right hand for the reference antenna. If PCAL (or something similar)
  is used the delay between left and right should be zero.
\item {\bf FRING} options: CALS 'target','calib'; APARM(3)=1 and APARM(5)=1\\
Finds the rates and combined phases for the target.
\item {\bf CALIB} options: CALS 'calib'; SOLMODE 'A\&P'; APARM(3)=1
and APARM(5)=1\\ Produces a well calibrated version of the target
which can be imaged. 
\item {\bf IMAGR}  Deconvolve this with, either a few model
components, or clean it and then box up the clean components into a
few regions (with CCEDT), and use it for the next stage.
\item {\bf LPCAL} options: CALS 'calib'; in2na 'calib' and in2c 'icln'\\
Does the polarisation calibration on the target, using the cleaned
model.
\item {\bf CALIB} options: CALS 'target'; SOLMODE 'A\&P'; DOPOL 2;
APARM(3)=1 and APARM(5)=1\\ 
Calibrates the target using the polarisation solutions.
\item {\bf IMAGR} Image the source in I, Q and U. The sum of the
  fluxes (total and polarised) should compare to the lower resolution
  (VLA or ATCA) values. The correction $\phi_{{\rm RL}}$ is
  $2\chi_{{\rm true}}-\arctan(\frac{\sum{\rm U}}{\sum{\rm Q}})$, for
  each IF.
\item {\bf CLCOR} options: CLCORP  $\phi_{{\rm RL}}$ stokes 'L' and OPC 'polr'\\
  Rotates the D-terms to match the calibration value of $\chi_{{\rm
      true}}$.
\end{itemize}

If this recipe is not followed the solutions for the L and R hand
polarisations are independent. This is not a problem if there is
sufficient signal to noise, however in mm-VLBI this is rarely the case
and the two polarisations need to be combined in the fringe search
stage. This is why it is important not to treat them as independent
except for the prime calibrator. 

\section{A list of the changes to classic AIPS}
\label{app:code}

A summary of the files changed:

\begin{itemize}
\item  LPCAL.FOR: Add a new subroutine LPCAL\_VIS
\item  LPCAL\_EXT.FOR: New program to allow dual D-terms
\item  PARANG.FOR: all mount types now supported
\item  PRTAN.FOR: Add mount names
\item  CLCOR.FOR: in ANAXIS allow all mount types
\item  DFCOR.FOR: in ANAXIS allow all mount types
\item  SNPLT.FOR: does not use PARANG, so mount types added
\item  DTSIM.FOR: (in ORIENT) does not use PARANG, so mount types added
\item  DTSIM.FOR: (in GETBAS) allow all mount types
\end{itemize}


\begin{thebibliography}{}

\bibitem{evn_78} Aaron, S., 1997, EVN Memo \#78,
``Calibration of VLBI polarisation data'', 
http://www.jive.nl/techinfo/evn\_docs/evn\_docs.html

\bibitem{brisken_evla} Brisken, W., 2003, eVLA Memo \#58, 
``Using Grasp8 To Study The VLA Beam'', 
http://www.aoc.nrao.edu/evla/geninfo/memoseries/evlamemo58.pdf

\bibitem{g339}
Dodson, R. 2007, A\&A, submitted

\bibitem{d03} Dodson, R., Lewis, D., McConnell, D., Deshpande, A.~A.\
2003.\ ``The radio nebula surrounding the Vela pulsar.'' Monthly
Notices of the Royal Astronomical Society 343, 116-124.

\bibitem{2000A&AS..143..515H} Hamaker, J.~P.\ 2000.\ ``Understanding
radio polarimetry. IV. The full-coherency analogue of scalar
self-calibration: Self-alignment, dynamic range and polarimetric
fidelity.'' Astronomy and Astrophysics Supplement Series 143, 515-534.
 
\bibitem{1995AJ....110.2479L} Leppanen, K.~J., Zensus, J.~A., Diamond,
P.~J.\ 1995. ``Linear Polarization Imaging with Very Long Baseline
Interferometry at High Frequencies.''Astronomical Journal 110, 2479.

\bibitem{1999ASPC..180..499K} Kemball, A.~J.\ 1999. ``VLBI
polarimetry.'' ASP Conf.~Ser.~180: Synthesis Imaging in Radio
Astronomy II, 180, 499.

\bibitem{ojha_lbap} Ojha,~R.,~2001, LBA~Workshop,\\
http://www.atnf.csiro.au/$\sim$rojha/lbapstuff/workhandout\_inc.ps

\bibitem{rioja_afpt} Rioja, M.~J., Dodson, R., Porcas, R.~W., Suda,
H., Colomer, F.\ 2005, ``Measurement of core-shifts with astrometric
multi-frequency calibration.'', 17th European VLBI Meeting for Geodesy
and Astometry, astro-ph/0505475.

\bibitem{TMS_book} Thompson, A.~R., Moran, J.~M., Swenson, G.~W.\
2001.\ ``Interferometry and Synthesis in Radio Astronomy'', 2nd
Edition. Wiley-Interscience
 
\end{thebibliography}
\end{document}